# Execution and Other Details for New Space Experiment Proposed for General Relativity

**Abhijit Biswas, Krishnan RS Mani** [*]

Indian Association for the Cultivation of Science, Calcutta 700 032, India

**Abstract**

Einstein stated in 1920 as one of the "Few Inferences from the General Theory of Relativity" that the special relativity postulate of constancy of speed of light, c, in vacuum, is not valid according to general relativity, as light rays could curve only when c varied with position. He had also stated the expression for the variable speed of light. However, only during the 1970's, it became conceivable to plan and conduct space experiment for determination of c at locations closer to sun and beyond the "spheres of gravitational influence" of the earth and other planets, where c may vary significantly, and thus can reveal the General relativistic nature of c, which relates to the variation of c under the influence of gravitational field. But, in 1983, for accuracy improvement of the length standard, the magnitude of c was frozen at a fixed value and the meter was redefined. This happened before any space experiment could be conducted to verify the effect of the gravitational influence on c, of massive celestial bodies like the sun, and no such experiment has been planned during more than three decades thereafter.

Now, a low-cost space experiment is being proposed based on a dictum from Einstein regarding the importance of continued experimentation for verifying the theoretical principles incorporated in his theory of relativity. For revealing the General relativistic nature of c, and for fulfilling a few other objectives, this direct Space experiment needs to be conducted for determining the magnitude of c by sending a miniaturized space qualified Laser device on board a spacecraft, during any of the future space missions to Mercury, Venus, or the sun.

It is expected that NASA, ESA, ASI, ISRO, or any other Space research organization would conduct this low-cost experiment, which is now possible even for any small country by sending one smaller than 100 kg micro-satellite, utilizing the launch-sharing facility of any other lending country.

[*] E-mail: godopy@vsnl.com

## 1. Introduction:

The speed of light, c, has intrigued scientists for more than three centuries and the concept of General relativistic nature of c is so counter-intuitive (as explained further below) that it continues to do so even now.

To develop better understanding of the relativistic nature of time, it had taken (as detailed below) almost seven decades of experimentation till the commissioning of Global Positioning System (GPS) in 1977. Similarly intriguing, the General relativistic nature of c, which relates to the variation of c under the influence of gravitational field, needs further experimentation for developing better comprehension.

Keeping the intriguing nature of relativistic principles in mind, Einstein [1] stated his first dictum in 1934, that the theory of relativity is a fine example of the fundamental character of the modern development of theoretical science; the initial hypotheses become steadily more remote and abstract from experience; on the other hand, it gets nearer to the grand aim of all science, which is to include the largest possible number of empirical facts by logical deduction from the least possible number of hypotheses or axioms.

For resolving the intriguing nature of relativistic principles, Einstein laid emphasis on the importance of continued experimentation for verifying the theoretical principles incorporated in his theory of relativity, and stated [2] specifically in his second dictum that it may equally well happen that clearly formulated principles lead to conclusions which fall entirely, or almost entirely, outside the sphere of reality at present accessible to our experience; and, in that case it may need many years of empirical research to ascertain whether the theoretical principles correspond with reality; and, we have an instance of this in the theory of relativity.

Einstein with great foresight and as a true scientist, dedicated the above-mentioned two-dictums-roadmap for future generations to make his general relativity theory (GRT) evolve to perfection, or in his own words, to take it nearer to the 'grand aim' of all science.

But, even after almost half a century of space-age experimentation on various aspects of light and relativity theory, no experiment has so far been conducted to determine c beyond the "sphere of gravitational influence" of the earth and other planets, and nearer to massive celestial bodies like the sun. Hence, the authors are proposing this low cost space experiment for verifying the prediction of Einstein's GRT, related to the effect of strong solar gravitational field on c, and for fulfilling a few other objectives.

## 2. Historical Background:

Maxwell made the speed of light, c, central to electromagnetism in the second half of nineteenth century. During the twentieth century, making a very bold attempt amongst his contemporary physicists, Einstein formulated the Special Relativity Theory (SRT) in 1905, by abandoning the ether theory, introducing the concept of relativistic time, and replacing Newton's three absolutes (space, time and mass) by only a single absolute, c, and establishing two paired-quantities, mass-energy and space-time, related through c.

But, the replacement of the concept of universal time with the concept of relativistic time is so counter-intuitive that it was used to be considered mysterious and shocking, till many experiments confirmed SRT during the decade after 1905. In 1915 while submitting his paper on GRT, Einstein used the expression "special theory of relativity" to distinguish it from GRT. Perhaps he avoided to mention about the influence of variations in gravitational field strength on c earlier (till SRT concepts became well founded from experimental verifications), so that the SRT concepts appear less mysterious or counter-intuitive. But, even after seven decades of successful experimentation on relativistic concepts, a section of physicists was unsure about the real nature of relativistic time during the GPS commissioning in 1977; this made Ashby state [3] that when the Cesium atomic clock was placed in orbit for the first time, it was acknowledged that orbiting clocks would need relativistic corrections, but there was uncertainty as to its sign as well as to its magnitude, and there were some who even disbelieved that relativistic effects were facts that would need to be included. It may be mentioned here that this situation persisted before GPS commissioning in spite of the fact that the high-precision macroscopic clock experiments by Hafele-Keating [4] and Vessot et al. [5] were successfully conducted at the beginning of the 1970's for developing better understanding of this relativistic effect. Ultimately, the relativistic nature of time could be better understood by the end of 1970's after GPS commissioning.

Similarly, the General relativistic nature of c, which relates to the variation of c under the influence of gravitational field needs further experimentation for better understanding, in view of a 'categorical statement' made by Einstein [6] in 1920 as one of the "Few Inferences from the General Theory of Relativity" that the law of the constancy of c in vacuum, which constituted one of the two fundamental postulates of the SRT, was not valid according to the GRT, as curvature of light rays can only occur when c varies with position. Einstein also stated [6] that SRT can not claim an unlimited domain of validity; and its results are valid only so long as it is possible to disregard the influences of gravitational fields on the phenomenon of light propagation.

### 3. The SRT postulate of constancy of the speed of light in vacuum

As SRT is simpler than GRT and can be understood more intuitively, by the time the would-be-physicists proceed beyond SRT to the learning stage of GRT, the SRT principle of constancy of c in vacuum become deeply embedded in the mind of even many renowned physicists, as an important and indispensable part of their relativistic concepts, as can be seen in the following example of Irwin Shapiro, who proposed in 1964 the famous Shapiro Time delay effect (which is known as the fourth test of GRT), more than four decades after Einstein himself proposed the first three tests.

C.M. Will mentions [7] that during his lecture in 1961, Irwin Shapiro was puzzled on hearing that c is not constant in GRT; at that time Shapiro, five years after his Ph.D. in physics from Harvard university was working at MIT's Lincoln laboratory, for improved determination of the AU (astronomical unit) by radar ranging to planets, and had already concluded that radar ranging could also be used to give improved measurements of the relativistic perihelion shift of Mercury's orbit. Will also mentioned [7] that Shapiro was perplexed because he had always thought that according to relativity, magnitude of c should be the same in every inertial frame; although he was aware that GRT predicts that light is deflected by a gravitating mass, he had the following unanswered question in his mind: Would the speed of light also be affected. Will further mentioned [7] how Shapiro utilized his corrected notion on the General relativistic nature of c for estimating the correct round-trip time of a radar signal to a distant object, and discovered the Shapiro Time delay effect, published in 1964. This is a real life example of a case in which eliminating an unproven or incorrect notion related to the General relativistic nature of c or the relativity theory, could be useful for improving one's relativistic concepts and making even new discoveries.

Almost a quarter century later, while reflecting on the contemporary knowledge amongst a section of scientists, Wilkie [8] mentioned in 1983 that if one reverses the logic of the experiments that physicists performed for determination of c during the 1970's, one can see that they verify Einstein's (SRT) assumption that c is constant, and show this assumption is correct within the present-day limits of experimental accuracy. But, this conclusion by Wilkie clearly does not conform to the above-mentioned 'categorical statement' made by Einstein.

Wilkie [8] further mentioned that these experiments were carried out at different times, in different places on the surface of the Earth, and at different points along the Earth's orbit round the Sun, and arrived at a misleading conclusion that c was found to remain constant within the present-day limits of experimental accuracy. The results of these experiments should be interpreted in the light of the continuation of the categorical statement by Einstein [6] that SRT can not claim an unlimited domain of validity; and its results are valid only when the influences of gravitational fields on the phenomenon of light propagation, is insignificant.

What Wilkie missed to take note of, is that at all the 'different locations on the surface of the Earth, and at different points along the Earth's orbit' as mentioned by him, the experiments were carried out within the "sphere of influence" of the gravitational field of the Earth (for example, three terrestrial laboratories L1, L2, L3, of three different countries, during three different times over a year, but always within the "Sphere of gravitational influence of Earth", are shown in Fig. 1 at Annexure I). At all those locations the gravitational field strength did not vary appreciably (as explained at the following paragraph) so as to affect the magnitude of c significantly and beyond its limits of experimental accuracy.

It may be noteworthy to mention here that the variation in the magnitude of c from that measured on the Earth surface to that measured at the terrestrial laboratories within the limit of the "sphere of influence" (having approximately a radius of 1,500,000 km) of Earth's gravitational field, is only a fraction of the limit of experimental accuracy (± 0.3 m/s) of the adopted 1983 value of c.

Thus, it is clear that Wilkie drew a misleading conclusion, which did not also conform to the above-mentioned 'categorical statement' made by Einstein. The right conclusion that Wilkie would have arrived at, if drawn in the light of the 'categorical statement' by Einstein [6], can be stated as**:** Since all those experiments were conducted within the "sphere of influence" of Earth's gravity, c was found to remain constant within the present-day limit of experimental accuracy; hence, for verifying the validity of the (SRT) principle of constancy of c, it is necessary to conduct a space experiment for determination of c at locations closer to sun, and adequately beyond the "spheres of gravitational influence" of the earth and other planets, where the stronger gravitational field of sun may significantly affect the measurement.

In view of the above, the authors are proposing this new space experiment that can reveal the General relativistic nature of c, which relates to the variation of c under the influence of gravitational field.

## 4. General relativistic nature of the Speed of Light

The General relativistic nature of c that relates to the variation of c under the influence of gravitational field, had been spelt out on several occasions by Einstein**:**

- In a 1911 paper [9], he expressed the equation for the variable speed as $c_r = c_0 \cdot (1 - F)$, where F = Red Shift Factor,
- In a 1912 paper [10], he stated that the principle of the constancy of the velocity of light, is valid only for space-time regions of constant gravitational potential.
- In his 1920 'categorical statement' [6] as one of the "Few Inferences from the General Theory of Relativity", he stated that the law of the constancy of c in vacuum, which constituted one of the two fundamental postulates of the SRT, was not valid according to the GRT, as curvature of light rays can only occur when c varies with position. Einstein further stated [6] that SRT can not claim an unlimited domain of validity; and its results are valid only so long as it is possible to disregard the influences of gravitational fields on the phenomenon of light propagation, and
- In his 1921 calculation of light deflection angle [11] also, he used an equation involving variable light velocity (L).

But, the fact that more than half a century after these statements by Einstein, Wilkie [8], had used a misleading 'reversed logic' in 1983 as mentioned in the previous section, shows that the concept of General relativistic nature of c is also as counter-intuitive as that of the relativistic nature of time, which needed more than seven decades, technological development of high-precision macroscopic clocks and three successive precision experiments (namely, Hafele-Keating [4], Vessot et al [5], and GPS application [3]) for better comprehension.

To develop better comprehension of the General relativistic nature of c, it was necessary to plan and conduct at least one space experiment beyond the "sphere of influence" of gravity of the Earth and other planets, for determining the effect of the strong solar gravitational field on the magnitude of c, after the cesium "clock" was adopted as the international standard in 1967, when it became possible to determine the second with an uncertainty of 1 part in $10^{13}$, and when it became conceivable to plan and conduct a space experiment for determination of c. But, during this era, the physicists faced a more stubborn problem for improving the measurement accuracy of the meter, the length standard, which was known to an accuracy level of only about 4 parts in $10^9$.

Wilkie [8] mentions that after the advent of the lasers in the optics laboratory, during the 1970's

- physicists could inter-compare the wavelength of laser beams to a few parts in $10^{11}$, but could not calculate the absolute wavelengths of their laser beams because the accuracy level of measuring a meter was poor;
- physicists had been determining c from the product of the measured frequency of a laser's radiation and its wavelength, and

- in almost all the cases of determination of the magnitude of c during the 1970's and the early 1980's, the quoted measurement uncertainties stemmed, not from any experimental limitation, but from the inaccuracy in measuring the meter.

Thus, the exigency of improving the measurement accuracy in the meter, prompted the international scientific community, through the Comité Consultatif pour la Définition du Mètre (CCDM), to a new definition for the meter [12], which was adopted by the 17th CGPM (Conférence Général des Poids et Mesures) in 1983. The last experimental value of c was reported as 299,792,458.6 ± 0.3 m/s, based on a 1983 laser measurement by NIST (formerly NBS), USA [13].

However, the 17th CGPM has neither labeled the adopted c, as the terrestrial c, nor called it a limiting speed. The 17th CGPM had just stated [12] that the adopted speed of light in vacuum is exactly 299792458 m/s.

For various reasons, the physicists have not felt any impetus to propose a space experiment during more than three decades next, for doing a direct and truly local (corresponding to the location of the spacecraft) measurement of **c** beyond the "sphere of gravitational influence" of the earth and other planets, and nearer to massive celestial bodies like the sun, though it is essential according to Einstein's second dictum on continued experimentation for verifying the theoretical principles incorporated in his relativity theory, and for developing a better comprehension of the General relativistic nature of c.

Hence, the authors are proposing this space experiment that may determine the value of c from a spacecraft location, just before it enters the "sphere of influence" of the gravitational field of Mercury or Venus, where the GRT-predicted magnitude of c (based on the above-mentioned Einsteinian equation for variable c) is respectively equal to 299792451 or 299792454 m/s, which are lower than the adopted standard value of c respectively by 7 or 4 m/s.

More than half a century before the advent of space-age, when it was inconceivable to do space-experiments for determination of c at locations beyond the "sphere of influence" of the gravitational field of the earth, Einstein had no option but to adopt the hypothesis that the terrestrial c is the limiting speed [14] of any object located anywhere in the universe, based on the prevalent knowledge of the contemporary physicists (viz., Lorentz, Poincare, etc.) as is evident from the following**:**
- Lorentz transformations which does not make sense when velocity of any object exceeds c, as presented by Lorentz in his 1904 paper titled "Electromagnetic Phenomena in a System Moving with A Velocity Less Than That of Light", and
- A statement by Poincare (as mentioned by Bernstein [15]) in 1904 that we should construct a totally new mechanics replacing the mechanics of Newton, according to which c would become an impassable limit, as inertia increases with the velocity of an object.

But, in connection with such hypothesis, Einstein's first dictum [1] states that the 'grand aim' of all science (including his theory of relativity), is to incorporate the largest possible number of empirical facts by logical deduction from the least possible number of hypotheses. Thus, in case, the outcome of conducting this experiment, following Einstein's second dictum mentioned earlier, on continued experimentation for verifying the theoretical principles incorporated in his relativity theory, leads to a new empirical fact – indicating that this hypothesis be replaced by a better one which may point towards a super-massive black hole at the center of any galaxy, the magnitude of c at whose proximity is equal to the upper limit of velocity in nature – it may alter Einstein's theory of Relativity to some extent, giving it a push, towards the 'grand aim' of perfection.

Anticipating such situations in applications of an abstruse theory like GRT, Einstein stated more than eight decades ago, these two dictums mentioned above, so that based on long-continued experimental verifications, more and more empirical facts can be incorporated while dropping the previously-adopted hypothesis or postulates, so that his relativity theory can evolve and get perfected, or in his own words, get nearer to the 'grand aim' of all science.

## 5. Discussion on the Proposed Experiment for the Speed of Light

The proposed experiment can be conducted from a solar probe or from a spacecraft (moving through the "sphere of influence" of the stronger gravitational field of the sun) before it enters the "sphere of influence" of the gravitational field of any one of the planets**:** Mercury or Venus.

The methodology of the proposed experiment can be a replication of the last one conducted in 1983 for visible light in a terrestrial laboratory, during a He-Ne laser measurement [13] by NBS (presently NIST), USA, employing the method of determination of laser frequency given at Jennings DA et al [16], or can be its modern version.

The terrestrial value of **c** as determined during the above-mentioned experiment in 1983 was 299792458.6 ± 0.3 m/s.

In view of the above, the proposed experiment should verify the magnitude of c at locations closer to the sun at an accuracy level of at least a fractional metre per second.

Recent technological advancements [17, 18] in very-compact robust optics have led to the development of miniaturized, space qualified lasers that are useful for such simple and low-cost space-borne experiments as the STAR (Space Time Asymmetry Research) Mission (a USA, Europe and SA collaboration), and also for the one proposed here.

For conducting this experiment, it will be necessary to send a miniaturized Iodine Stabilized Helium-Neon (633 nm) laser device along with its controller unit, on board a spacecraft.

The advent of miniaturized, fully space qualified laser systems has not only brought down the cost of conducting such space experiments, but also has enabled even the small technologically aspiring countries to conduct such space experiments, as at present

- micro-satellites (weighing less than 100 kg) with smaller payloads can be launched from a few other country's launch-sharing facility, and
- micro-satellite bus are commercially available as modular, mass efficient structure, provided with
    - 24-36 V Electrical Power subsystem,
    - Telemetry, tracking and Command subsystem,
    - Command and data handling subsystem, having data bus with standard 1553 interfaces, and flight software resident on the flight processor,
    - Attitude/velocity Determination and Control Subsystem,
    - Thermal-control subsystem, and
    - Propulsion subsystem.

The outcome of the proposed test would however not necessitate any immediate change in the already adopted value of c, or in the adopted definition of the meter by the 17th CGPM in 1983, even if the value obtained from the proposed experiment is found to be different from the GRT-predicted value of c. Because, the 1983-adopted value of c is really the value of c determined from terrestrial experiments that correspond to the influence of the strength of Earth's gravitational field at terrestrial laboratories. Thus, when a typical miniaturized Laser system calibrated in a terrestrial laboratory is sent to a space mission (say, bound for Mercury or Venus), it should indicate a lower value of laser frequency under the influence of the strong solar gravitational field; and, this lower value of laser frequency when multiplied by the appropriate laser wavelength, should give the GRT-derived local value (corresponding to the spacecraft location) of c.

6. **Conclusion**

Following Einstein's second dictum on continued empirical research for checking whether the theoretical principles incorporated in his theory of relativity correspond with reality, this low-cost Space experiment needs to be conducted for fulfilling the following objectives:

- To provide the experimental verification of an important prediction of Einstein's general relativity regarding the General relativistic nature of c, which relates to the variation of c under the influence of gravitational field, which is long overdue.
- To provide a first ever direct and truly local measurement of **c** beyond the "sphere of influence" of earth's gravity, and close to a massive celestial body like the sun (that is, at locations under the influence of the stronger solar gravitational field).
- To directionally indicate whether the magnitude of c, measured from terrestrial experiment sets the upper limit of velocity for any material body located anywhere in the universe.
- To deepen every relativist's comprehension of relativity theory, gravitation and relativistic effects, and

- To lead consequently, perhaps to new direction for theoretical physicists and to new perspectives for unifying gravity with the other three fundamental interactions.

Space research organizations of various countries either jointly or singly, have either conducted, or are conducting or planning for, many costly (and sometimes long-term) experiments viz., GP-B, LIGO, STEP, LISA, STAR etc.

Now, it is expected that any of the Space research organizations (viz., NASA, ESA, ASI, ISRO, etc.) would conduct this low-cost experiment, during Space missions to Mercury, Venus, or the sun, or may choose to conduct it as an independent space mission as the cost may be brought down by sending micro-satellites, weighing less than 100 kg, using the launch-sharing facility of any other lending country.

To utilize the opportunity to conduct such a low-cost experiment following Einstein's dictum on continued empirical research for verifying the theoretical principles incorporated in his theory of relativity, and make a historic contribution towards the evolution of Einstein's theory of Relativity, at present, even a small country, not having a Space research organization, (or, even an advanced country desirous of conducting this Space experiment at low-cost), may contract with an experienced organization, or hire the services of an experienced project director who

- can organize the planning, procurement, and launching of a micro satellite carrying the laser device, from another country's launch pad (adopting the piggy-back launching method), and
- can guide (else, can hire the services of a Principal Investigator, from BIPM {France}, Stanford Univ., JILA, NIST, {USA}, or similar other institutes) remote operation of the laser device for a few days.

## 7. Acknowledgements:

The authors gratefully acknowledge the help provided to them by IACS (Indian Association for the Cultivation of Science, Calcutta 700 032, India) and Director, JPL (Jet Propulsion Laboratory, Caltech, California, USA) during the early stages of more than two decades long persevering work that led to the development of their work on relativity theory. In particular, it must be mentioned that over several years Prof. N.C. Sil, Head of the Dept. of Theoretical Physics, IACS, and Dr. E. Myles Standish, Jr., head of Ephemeris Division at JPL, provided much guidance and help.

## 8. References:

[1] A. Einstein, "The Problem of Space, Ether, and the Field in Physics", *Mein Weltbild* (Querido Verlag, Amsterdam, 1934), *Ideas and Opinions* by Albert Einstein, Ed. C. Seeling, Tr. S. Bargmann, (Crown Publishers, New York, 1954), p. 282 (p.289 in PDF).
https://namnews.files.wordpress.com/2012/04/29289146-ideas-and-opinions-by-albert-einstein.pdf

[2] A. Einstein, "Principles of Theoretical Physics", *Mein Weltbild* (Querido Verlag, Amsterdam, 1934), *Ideas and Opinions* by Albert Einstein, Ed. C. Seeling, Tr. S. Bargmann, (Crown Publishers, New

York, 1954), p. 222 (p.229 in PDF). https://namnews.files.wordpress.com/2012/04/29289146-ideas-and-opinions-by-albert-einstein.pdf

[3] N. Ashby, Living Reviews in Relativity, http://relativity.livingreviews.org/Articles/lrr-2003-1/

[4] J.C. Hafele and R.E. Keating, Science **177**, 168 (1972).

[5]. R.F.C. Vessot, L.W. Levine, E.M. Mattison, E.L. Blomberg, T.E. Hoffman, G.U. Nystrom, B.F. Farrel, R. Decher, P.B. Eby, C.R. Baugher, J.W. Watt, D.L. Teuber, and F.O. Wills, Phys. Rev. Lett..**45**, (1980) p. 2081.

[6] A. Einstein, *Relativity: The Special and General Theory* (Methuen & Co., London, 1920), Chap. XXII, http://lipn.univ-paris13.fr/~duchamp/Books&more/Penrose/%5BAlbert_Einstein,_Roger_Penrose,_Robert_Geroch,_Da.pdf (p.97 of Book, or p.126 of PDF).

[7] C. M. Will: *Was Einstein Right? Putting General Relativity to the Test,* Oxford University Press, Oxford, 1988.

[8] T. Wilkie: "Time to Remeasure the Metre", *New Scientist*, **100**, No. 1381, (1983) pp. 258-263,

[9] A. Einstein, "On the influence of Gravitation on the Propagation of Light", *Annalen der Physik*, Vol.35, pp. 898-908, 1911, English Tr., Equation at p.903 (p.7 in PDF)
https://www.relativitycalculator.com/pdfs/On_the_influence_of_Gravitation_on_the_Propagation_of_Light_English.pdf

[10] A. Einstein, "The Speed of Light and the Statics of the Gravitational Field", *Annalen der Physik,* 38 (1912): 355-69]; Eng.Translation at p.95 (http://einsteinpapers.press.princeton.edu/vol4-trans/107?ajax)**:**

[11] A. Einstein, *The Meaning of Relativity,* (New York: Barns & Noble).
https://www.gutenberg.org/files/36276/36276-pdf.pdf

[12] B.N. Taylor (Ed.)**:** "The International System of Units (SI)", NIST Special Publication 330, Edn. 2001, p.37. http://physics.nist.gov/Pubs/SP330/sp330.pdf

[13] Jennings DA et al., The Continuity of the Meter**:** The Redefinition of the Meter and the Speed of Visible Light, *Journal of Research of the National Bureau of Standards*, Vol.92 (1987), pp.11-12. http://nvlpubs.nist.gov/nistpubs/jres/092/1/V92-1.pdf

[14] A. Einstein and L. Infeld, *The Evolution of Physics* (Cambridge University Press, Cambridge, 1938).
https://archive.org/details/evolutionofphysi033254mbp , p. 204 (p.224 in PDF).

[15] J. Bernstein, *Einstein* (Penguin Books, 1976).

[16] Jennings DA et al., Direct frequency measurement of the $I_2$-stabilized He–Ne 473-THz (633-nm) laser, *Optics Letters*, Vol. 8, Issue 3, pp. 136-138 (1983) http://dx.doi.org/10.1364/OL.8.000136.

[17] Robert Byer et al, Space Time Asymmetry Research (STAR),
http://www.tapir.caltech.edu/~kamion/CFP/Byer_Space_Time_Asymmetry_Research_CFP.pdf (accessed Sept. 2015)

[18] Ke-Xun Sun et al, Technology Development for Space Time Asymmetry Research (STAR) Mission,
https://www.researchgate.net/publication/255667067_Technology_Development_for_Space_Time_Asymmetry_Research_(STAR)_Mission www8.nationalacademies.org/astro2010/detailfiledisplay.aspx?id=531 (accessed Sept. 2015)

# Hyperbolic trajectory or orbit of a photon during Shapiro time delay experiment and Locations of three terrestrial laboratories

**Legend:** L1, L2, L3, are locations of terrestrial laboratories of three different countries, shown during three different times over a year, but always within the "Sphere of gravitational influence of Earth".
Color zones depict the individual "Sphere of gravitational Influence":

- ☐ Gray of the Sun;
- ☐ Light Blue of the Earth;
- ☐ Light Yellow of the Mars.

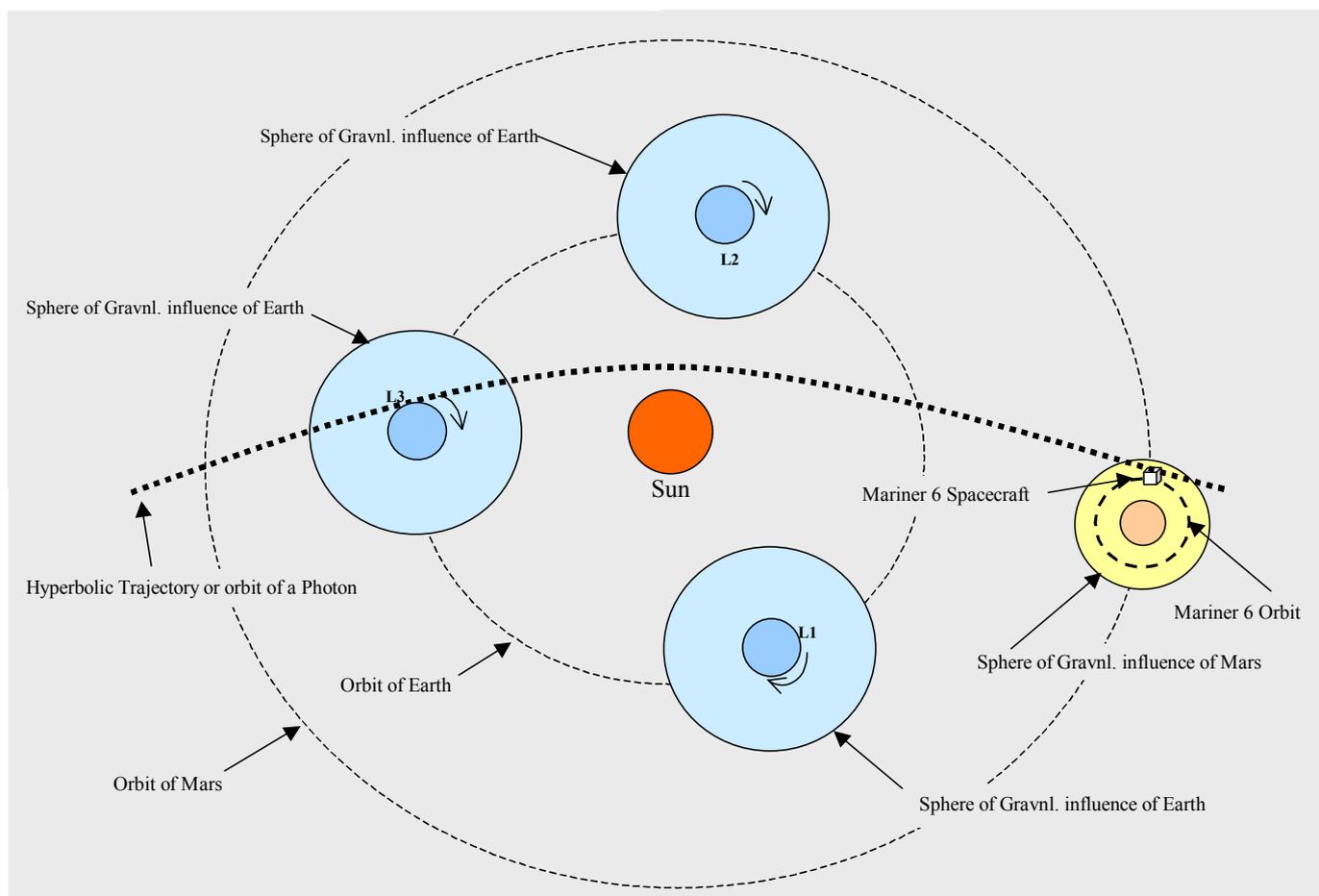

..
**Fig. 1.** Hyperbolic trajectory or orbit of a photon (released from Mariner 6 spacecraft during Shapiro time delay experiment) in the heliocentric coordinate system, and
Locations of three laboratories L1, L2, L3, of three different countries, shown during three different times over a year, where experiments were conducted for determination of magnitude of c.